\documentclass[aps,prd,english,preprint,superscriptaddress]{revtex4}
\usepackage{epsfig,epsf,amsmath,amsthm,amsfonts,
            graphics,graphicx,amssymb,babel}
\newcommand{\Dzero}{D\O\ }
\newcommand{\mll}{M_{\ell \ell}}
\newcommand{\ppbar}{$p\overline{p}$~}

\def\simge{\mathrel{%
   \rlap{\raise 0.511ex \hbox{$>$}}{\lower 0.511ex \hbox{$\sim$}}}}
\def\simle{\mathrel{
   \rlap{\raise 0.511ex \hbox{$<$}}{\lower 0.511ex \hbox{$\sim$}}}}
\def\lessim{\mathrel {\vcenter {\baselineskip 0pt \kern 0pt
\hbox{$<$} \kern 0pt \hbox{$\sim$} }}}
\def\gessim{\mathrel {\vcenter {\baselineskip 0pt \kern 0pt
\hbox{$>$} \kern 0pt \hbox{$\sim$} }}}
\def \rightdownarrow
 {\kern.4em {\raise1.75ex\hbox{$\Bigl |$}$\kern-0.27em{\longrightarrow}$}}
\def \mrightdownarrow
 {\kern.4em {\raise1.25ex \hbox{$|$} \kern-0.27em{
                                                \longrightarrow}}}

\def\mMEt{\not\kern-.35em {E_T}}

\def\met{\mbox{${\hbox{$E$\kern-0.6em\lower-.1ex\hbox{/}}}_T$}} 

\begin{document}
%
\title{Measurement of the $b$ Jet Cross Section in Events with a $Z$ Boson
in \ppbar Collisions at $\sqrt{s}$=1.96~TeV
}

\affiliation{Institute of Physics, Academia Sinica, Taipei, Taiwan 11529, Republic of China} 
\affiliation{Argonne National Laboratory, Argonne, Illinois 60439, USA} 
\affiliation{Institut de Fisica d'Altes Energies, Universitat Autonoma de Barcelona, E-08193, Bellaterra (Barcelona), Spain} 
\affiliation{Baylor University, Waco, Texas  76798, USA} 
\affiliation{Istituto Nazionale di Fisica Nucleare, University of Bologna, I-40127 Bologna, Italy} 
\affiliation{Brandeis University, Waltham, Massachusetts 02254, USA} 
\affiliation{University of California, Davis, Davis, California  95616, USA} 
\affiliation{University of California, Los Angeles, Los Angeles, California  90024, USA} 
\affiliation{University of California, San Diego, La Jolla, California  92093, USA} 
\affiliation{University of California, Santa Barbara, Santa Barbara, California 93106, USA} 
\affiliation{Instituto de Fisica de Cantabria, CSIC-University of Cantabria, 39005 Santander, Spain} 
\affiliation{Carnegie Mellon University, Pittsburgh, PA  15213, USA} 
\affiliation{Enrico Fermi Institute, University of Chicago, Chicago, Illinois 60637, USA} 
\affiliation{Joint Institute for Nuclear Research, RU-141980 Dubna, Russia} 
\affiliation{Duke University, Durham, North Carolina  27708, USA} 
\affiliation{Fermi National Accelerator Laboratory, Batavia, Illinois 60510, USA} 
\affiliation{University of Florida, Gainesville, Florida  32611, USA} 
\affiliation{Laboratori Nazionali di Frascati, Istituto Nazionale di Fisica Nucleare, I-00044 Frascati, Italy} 
\affiliation{University of Geneva, CH-1211 Geneva 4, Switzerland} 
\affiliation{Glasgow University, Glasgow G12 8QQ, United Kingdom} 
\affiliation{Harvard University, Cambridge, Massachusetts 02138, USA} 
\affiliation{Division of High Energy Physics, Department of Physics, University of Helsinki and Helsinki Institute of Physics, FIN-00014, Helsinki, Finland} 
\affiliation{University of Illinois, Urbana, Illinois 61801, USA} 
\affiliation{The Johns Hopkins University, Baltimore, Maryland 21218, USA} 
\affiliation{Institut f\"{u}r Experimentelle Kernphysik, Universit\"{a}t Karlsruhe, 76128 Karlsruhe, Germany} 
\affiliation{High Energy Accelerator Research Organization (KEK), Tsukuba, Ibaraki 305, Japan} 
\affiliation{Center for High Energy Physics: Kyungpook National University, Taegu 702-701, Korea; Seoul National University, Seoul 151-742, Korea; and SungKyunKwan University, Suwon 440-746, Korea} 
\affiliation{Ernest Orlando Lawrence Berkeley National Laboratory, Berkeley, California 94720, USA} 
\affiliation{University of Liverpool, Liverpool L69 7ZE, United Kingdom} 
\affiliation{University College London, London WC1E 6BT, United Kingdom} 
\affiliation{Centro de Investigaciones Energeticas Medioambientales y Tecnologicas, E-28040 Madrid, Spain} 
\affiliation{Massachusetts Institute of Technology, Cambridge, Massachusetts  02139, USA} 
\affiliation{Institute of Particle Physics: McGill University, Montr\'{e}al, Canada H3A~2T8; and University of Toronto, Toronto, Canada M5S~1A7} 
\affiliation{University of Michigan, Ann Arbor, Michigan 48109, USA} 
\affiliation{Michigan State University, East Lansing, Michigan  48824, USA} 
\affiliation{Institution for Theoretical and Experimental Physics, ITEP, Moscow 117259, Russia} 
\affiliation{University of New Mexico, Albuquerque, New Mexico 87131, USA} 
\affiliation{Northwestern University, Evanston, Illinois  60208, USA} 
\affiliation{The Ohio State University, Columbus, Ohio  43210, USA} 
\affiliation{Okayama University, Okayama 700-8530, Japan} 
\affiliation{Osaka City University, Osaka 588, Japan} 
\affiliation{University of Oxford, Oxford OX1 3RH, United Kingdom} 
\affiliation{University of Padova, Istituto Nazionale di Fisica Nucleare, Sezione di Padova-Trento, I-35131 Padova, Italy} 
\affiliation{LPNHE, Universite Pierre et Marie Curie/IN2P3-CNRS, UMR7585, Paris, F-75252 France} 
\affiliation{University of Pennsylvania, Philadelphia, Pennsylvania 19104, USA} 
\affiliation{Istituto Nazionale di Fisica Nucleare Pisa, Universities of Pisa, Siena and Scuola Normale Superiore, I-56127 Pisa, Italy} 
\affiliation{University of Pittsburgh, Pittsburgh, Pennsylvania 15260, USA} 
\affiliation{Purdue University, West Lafayette, Indiana 47907, USA} 
\affiliation{University of Rochester, Rochester, New York 14627, USA} 
\affiliation{The Rockefeller University, New York, New York 10021, USA} 
\affiliation{Istituto Nazionale di Fisica Nucleare, Sezione di Roma 1, University of Rome ``La Sapienza," I-00185 Roma, Italy} 
\affiliation{Rutgers University, Piscataway, New Jersey 08855, USA} 
\affiliation{Texas A\&M University, College Station, Texas 77843, USA} 
\affiliation{Istituto Nazionale di Fisica Nucleare, University of Trieste/\ Udine, Italy} 
\affiliation{University of Tsukuba, Tsukuba, Ibaraki 305, Japan} 
\affiliation{Tufts University, Medford, Massachusetts 02155, USA} 
\affiliation{Waseda University, Tokyo 169, Japan} 
\affiliation{Wayne State University, Detroit, Michigan  48201, USA} 
\affiliation{University of Wisconsin, Madison, Wisconsin 53706, USA} 
\affiliation{Yale University, New Haven, Connecticut 06520, USA} 
\author{A.~Abulencia}
\affiliation{University of Illinois, Urbana, Illinois 61801, USA}
\author{D.~Acosta}
\affiliation{University of Florida, Gainesville, Florida  32611, USA}
\author{J.~Adelman}
\affiliation{Enrico Fermi Institute, University of Chicago, Chicago, Illinois 60637, USA}
\author{T.~Affolder}
\affiliation{University of California, Santa Barbara, Santa Barbara, California 93106, USA}
\author{T.~Akimoto}
\affiliation{University of Tsukuba, Tsukuba, Ibaraki 305, Japan}
\author{M.G.~Albrow}
\affiliation{Fermi National Accelerator Laboratory, Batavia, Illinois 60510, USA}
\author{D.~Ambrose}
\affiliation{Fermi National Accelerator Laboratory, Batavia, Illinois 60510, USA}
\author{S.~Amerio}
\affiliation{University of Padova, Istituto Nazionale di Fisica Nucleare, Sezione di Padova-Trento, I-35131 Padova, Italy}
\author{D.~Amidei}
\affiliation{University of Michigan, Ann Arbor, Michigan 48109, USA}
\author{A.~Anastassov}
\affiliation{Rutgers University, Piscataway, New Jersey 08855, USA}
% ...
\collaboration{CDF Collaboration}
\noaffiliation

%
%

\begin{abstract}
\noindent A measurement of the inclusive bottom jet cross
section is presented for events containing a $Z$ boson in $p\bar{p}$
collisions at $\sqrt{s}=1.96$~TeV using the Collider Detector at Fermilab. 
$Z$ bosons
are identified in their electron and muon decay modes, and $b$ jets
with $E_T>20$~GeV and $|\eta|<1.5$ are identified by reconstructing a
secondary decay vertex. The measurement is based on an integrated
luminosity of about $330$~${\rm pb}^{-1}$. A cross
section times branching ratio of 
$\sigma (Z+b {\,\rm jets}) \times {\cal B}(Z \to \ell^+ \ell^-)= 0.93 \pm
0.36$~pb is found, where ${\cal B}(Z\to \ell^+ \ell^-)$ is the branching ratio of
the $Z$ boson or $\gamma^*$ into a  single flavor dilepton pair 
($e$ or $\mu$)   
in the mass range between $66$  and $116$ GeV$/c^2$.  The
ratio of $b$ jets to the total number of jets of any flavor in the $Z$
sample, within the same kinematic range as the $b$ jets, is
 $2.36 \pm 0.92\%$. Here, the  uncertainties are the quadratic sum of
statistical and systematic uncertainties. Predictions made 
with NLO QCD agree, within experimental and theoretical uncertainties, with these measurements.

\end{abstract}
\maketitle

\newpage
\section{Introduction}
The measurement of the $Z + b$ jet production cross section provides an important 
test of quantum-chromodynamics (QCD) calculations~\cite{zbmcfm}. The cross section  is sensitive to the $b$ quark density in the proton
and thus tests the perturbative calculations of this quantity. 
A precise knowledge of the $b$ quark density is
essential to accurately predict the production 
of particles that couple strongly to $b$ quarks including Higgs bosons ($h$) 
within supersymmetry models ($gb \to hb$, $bb \to h$) ~\cite{bbh,mssmhiggs} or
single top production~\cite{singletop} within the standard model 
($qb \to q^\prime t$ and $gb \to Wt$). The  $Z + b$ jet  cross section is also an important
test of the background predictions to standard model Higgs boson
production in association with a $Z$ boson, $ZH \rightarrow Zb\bar{b}$~\cite{smhiggs}.

The $Z$ cross section~\cite{wzinclprl, wzinclprd, d0wz} and $Z$+jets
cross section~\cite{zjets} have been measured at the Tevatron. Next to
leading order (NLO) QCD calculations are found to describe the
data. In this paper,  the first measurement of the $b$ jet cross
section for events with a $Z$ boson using the Collider
Detector at Fermilab (CDF)~\cite{CDF_run2} is reported. A similar measurement has been
made recently by the \Dzero collaboration~\cite{zbjetd0}.

The dominant production diagrams are $g b \rightarrow Zb$ and
$q\bar{q} \rightarrow Z b \bar{b}$: in NLO calculations they
contribute about 65\% and 35\%, respectively. At present the $b$ quark
density is derived from the gluon distribution function~\cite{collins} and
agrees well with the available measurements of the contribution to the
proton structure function $F_2$ for $Q^2<1,000$~GeV$^2$~\cite{h1},
where $Q^2$ is the momentum transfer squared.  The  measurement reported
in this paper is
sensitive to parton densities at higher values with $Q^2$
approximately equal to the square of the $Z$ mass ($M_Z$).

The analysis uses Run II $p\bar{p}$ collision data from CDF taken 
up to September 2004 at a center of mass energy of $\sqrt{s}=1.96$~TeV. The 
measurement is made by searching for pairs of electrons or muons with 
an invariant mass consistent with $M_Z$  and jets which contain 
a displaced secondary vertex consistent with the decay of a 
long-lived bottom hadron.  The light ($u$, $d$, $s$, and gluon) and charm ($c$) 
jets remaining after this vertex requirement are distinguished 
from the $b$ jets using the mass distribution of the charged 
particles forming the secondary vertex. This technique exploits 
the larger mass of the $b$ quark compared with light and $c$ quarks. 
The $Z$ cross section is defined
to include the irreducible Drell-Yan contribution $\gamma^* \rightarrow \ell^+\ell^-$
within the dilepton invariant mass range $66<\mll<116$~GeV/$c^2$. Note that this
cross section is numerically only $0.4\%$ higher than the inclusive $Z$ cross section 
independent of the mass range~\cite{wzinclprl}. The $Z + b$ jet production cross section 
is defined to be proportional to
the number of $b$ jets with jet tranverse energy $E^{\rm jet}_T>20$~GeV and pseudorapidity $|\eta^{\rm jet}|<1.5$ contained
in  events with a $Z$ boson.

In Sec. \ref{sec:detector} a brief description of the CDF detector
is given, and in Sec.~\ref{sec:mc} the Monte Carlo simulation is
described. Section
\ref{sec:selection} summarizes the event selection and the background
sources. In Sec.~\ref{sec:bfraction} the fraction of $b$ jets
within the data sample is determined.  In Sec.~\ref{sec:crosssection} 
the method to measure the cross section is
described, and in Sec.~\ref{sec:sys} the sources of systematic 
uncertainties are discussed. The
results of the measurement are given in Sec.~\ref{sec:result}, 
and a conclusion is presented in Section
\ref{sec:conclusion}.

\section{The CDF II Detector}
\label{sec:detector}
The CDF II detector is described in detail
elsewhere~\cite{CDF_run2}. It is a general purpose, nearly hermetic
detector situated around the $p\bar{p}$ collision point. A coordinate
system is used, in which $\theta$ is the polar angle with respect to
the proton beam direction, $\phi$ is the azimuthal angle and
$\eta=-\ln \tan(\theta/2)$ is the pseudorapidity. The transverse
energy and transverse momentum of a particle is defined as
$E_T=E\sin\theta$ and $p_T = p \sin \theta$, respectively, where $E$
is the energy measured by the calorimeter and $p$ is the momentum
measured in the tracking system. The missing transverse energy vector
is defined as $\vec{\met}=-\sum_i E_T^i \vec{n_i}$, where $\vec{n_i}$
is a unit vector that points from the interaction vertex to the center
of the $i$th calorimeter tower in the transverse plane and $E_T^i$ is
the transverse energy of the $i$th tower. The quantity $\met$ is the
magnitude of $\vec{\met}$, which is corrected for all identified muons
in an event \cite{wzinclprd}.

The transverse momenta of charged particles are measured by an eight-layer
silicon strip detector~\cite{L00,SVX,ISL} and the central outer
tracker (COT), a 96-layer drift chamber~\cite{COT} located inside a
solenoid that provides a 1.4~T magnetic field.  The innermost
layer of the silicon detector is located on the beryllium beampipe at
a radius of $1.5$ cm, and the outermost layer is located at $28$ cm.
The silicon detector provides tracking in the pseudorapidity region
$|\eta|<2$, with partial coverage up to $|\eta|<2.8$. The single hit resolution
is about 11~$\mu$m. Located outside 
of the silicon detector, the COT is a $3.1$~m long, open-cell drift
chamber with an active tracking region extending radially from $41$ cm
to $137$ cm. The COT provides coverage for $|\eta|<1$. For 
tracks with $p_T>1.5$ GeV/$c$  and
silicon hits the resolution on the impact parameter is about $34$~$\mu$m~\cite{cdfjpsi}, including the transverse size of the beam of about $25$~$\mu$m.

Located outside the solenoid, a segmented sampling calorimeter is used
for the measurement of particle energies. The central part of the
calorimeter covers the region $|\eta|<1.1~$\cite{cem,cha}, and the forward 
part of the calorimeter consists of two identical detectors covering 
$1.1 <|\eta|<3.6$~\cite{pem}. The central calorimeter uses
lead-scintillator sampling in the electromagnetic compartment and 
steel-scintillator sampling in the hadronic compartment. It is
instrumented with proportional strip and wire chambers [central
electromagnetic shower maximum detector (CES)]. They are located at a
depth of about six radiation lengths where the lateral profile of
electromagnetic showers is expected to be maximal and have a
segmentation of $1.5$~cm. The forward calorimeter uses 
lead-scintillator sampling for the
electromagnetic compartment and iron-scintillator for the
hadronic compartment. Further details about the calorimeters can be
found in Ref.~\cite{jes_nim}.

Drift chambers, located outside the central
hadron calorimeters and  behind  $60$~cm of iron shielding,
detect muons with $|\eta|<0.6$~\cite{cmu}.  Additional drift chambers
and scintillation counters detect muons in the regions
$0.6<|\eta|<1.0$ and $1.0<|\eta|<1.5$.  
Gas Cherenkov counters ~\cite{CLC} measure the
average number of $p\bar{p}$ inelastic collisions per bunch crossing
to determine the  luminosity.
\section{Monte Carlo Simulation}
\label{sec:mc}
A Monte Carlo  simulation is used to correct for inefficiencies due to
the selection requirements and detector effects. The Monte Carlo
generator {\sc pythia} v6.2 \cite{pythia} is used to generate the Drell-Yan signal
and the background processes, using CTEQ5L parton density functions
(PDFs)~\cite{cteq5l}. An underlying event model that describes the interactions of the
spectator partons and initial state QCD radiation has been included in
the generation. This model has been tuned to describe the Tevatron
data~\cite{jes_nim,ue}. 
The decays of the $b$ hadrons are generated by
the Monte Carlo generator {\sc QQ} v9.1~\cite{qq}. The CDF detector
response is simulated using a {\sc geant} based detailed
detector simulation \cite{geant,cdfsim}. 

The Drell-Yan Monte Carlo samples are normalized to
the next to next to leading order (NNLO) QCD cross section of $251.3$~pb
\cite{znnlo} for $66<\mll<116$~GeV/$c^2$, so that comparisons with the
data can be made.
 For $t\bar{t}$ and $ZZ$ processes, which contribute to the background, the NLO QCD cross
sections are used for the normalization: $\sigma_{t\bar{t}}=6.77$~pb~\cite{ttbar} 
and $\sigma_{ZZ}=1.4$~pb~\cite{zz}. 

Simulated events are reconstructed in the same manner
as the data events, and  the same event selection criteria are applied.

\section{Event Selection and Background} 
\label{sec:selection}

\subsection{Event Selection}
$Z$ bosons are detected in their decays into two electrons or two muons
with an invariant mass of the two leptons $\mll$ ($\ell=e$ or $\mu$)
between $66$ and
$116$~GeV/$c^2$.  The trigger requirements and the lepton selection 
follow closely those described in detail in Ref.~\cite{wzinclprd}.

Electrons are triggered by requiring a cluster of electromagnetic
energy with $E_T>18$~GeV and $|\eta|<1.1$ matched to a track with
$p_T>10$~GeV/$c$.  Further requirements are made on  position matching
and the shower shape in the CES. At least one trigger electron candidate is required. The
second electron candidate can either be in the central or forward
calorimeter, and looser identification criteria are imposed.  For
forward electron candidates no matching track is required. All
electron candidates are required to be isolated from other calorimeter
energy deposits~\cite{wzinclprd}.

Muons are triggered by requiring a track with $p_T>18$~GeV/$c$ and 
$|\eta|<1.0$ and a
track segment in the muon chambers that matches the extrapolated
position of the  track. At least one muon candidate that 
satisfies the trigger requirements is required. 
The other muon candidate is not required to have  signals in the muon chambers. 
All muon candidates are required to have a 
calorimeter energy deposit consistent with that of a muon 
and to be isolated from
other energy depositions~\cite{wzinclprl,wzinclprd}. The two highest $p_T$ muon candidates
are required to have opposite electric charges. 

Candidate $Z$ boson events are selected if $66<\mll<116$~GeV/$c^2$. A total of  27,659 candidate 
events  are observed in the electron channel and 15,698 events in the muon channel.

Having selected an event with a $Z$ boson candidate,  jets with 
 $E^{\rm jet}_T>20$~GeV and 
$|\eta^{\rm jet}|<1.5$ are searched for. Jets are defined by a cone jet algorithm with
a cone size $R=
\sqrt{(\Delta\eta)^2+(\Delta \phi)^2}=0.7$~\cite{jes_nim}. The jet energy is
 corrected to the hadron level energy. The hadron level energy is
defined to include  all
 particles from the $p\bar{p}$ collision within the jet cone,
 including particles from the hard scatter, multiple parton-parton
 interactions, and beam remnants. The jet energy is also corrected for 
particles produced in additional $p\bar{p}$ interactions, reconstructed in
the same bunch crossing. The jets are not corrected to the parton level
to be independent of the Monte Carlo modeling of this correction.

Events must satisfy either $\met<25$~${\rm GeV}$ or $H_T<150$~${\rm GeV}$,
where $H_T$ is the scalar sum of $\met$ and 
the transverse energies of all leptons and
jets in the event~\cite{ht}. This requirement reduces background from $t\bar{t}$ 
events in which the $t$ decays to $Wb$ and both the $W$ bosons decay
leptonically ($W \rightarrow l \nu$)
 by about $80 \%$, while reducing the signal by only $4 \%$,
as determined from Monte Carlo simulation.

A $b$ jet is defined as any jet that has at least one $b$ hadron within a
 cone of 0.7 around the jet axis. In this analysis a $b$ jet is identified
through the presence of a displaced vertex within the jet arising from
the decay of the long-lived bottom hadron. The algorithm used
was optimized for the measurement of the top quark production cross section~\cite{secvtx} 
but found to give adequate efficiency and purity for the present analysis.
A jet that has a reconstructed displaced vertex is called a ``$b$ tagged''
jet. The displaced vertex algorithm uses a two-pass approach to find a
secondary vertex. In the first pass an attempt is made to reconstruct
a secondary vertex using one track with $p_T>1.0$~${\rm GeV}/c$ and 
 two or more additional tracks with $p_T>0.5$~${\rm GeV}/c$, and all
tracks are required to have an impact parameter significance
$d_0/\sigma_{d_0}>2.5$. Here, $d_0$ is the minimum distance between
the track and the primary vertex in the plane transverse to the beam
direction and has uncertainty $\sigma_{d_0}$. If the first pass is
unsuccessful, a second pass is made using two tracks with
$p_T>1.5$~${\rm GeV}/c$ for one track, $p_T>1.0$~${\rm GeV}/c$ for the
other and $d_0/\sigma_{d_0}>3$ for both. The jet is labeled as a tag
if the transverse displacement significance
$|L_{2D}|/\sigma_{L_{2D}}>7.5$. Here, $L_{2D}$ is the distance from the
primary vertex to the secondary vertex in the plane transverse to the
beam direction projected onto the jet axis, and $\sigma_{L_{2D}}$ is
the estimated uncertainty.  The distance $L_{2D}$ is defined as
positive if the angle between the transverse displacement and the jet
direction is less than $90^\circ$, and as negative otherwise. A {\it
positive tag} has positive $L_{2D}$, and a {\it negative tag} has
negative $L_{2D}$.

The $b$ tagging efficiency is determined as a function of  $E^{\rm jet}_T$
and $\eta^{\rm jet}$ from a separate data sample of about 28,000 dijet events,
where one of the jets has a reconstructed semileptonic $b$ or $c$
decay~\cite{secvtx}. The ratio of the efficiency in data to that in
Monte Carlo simulation is found to be 
$0.91 \pm 0.06$. The average
data $b$ tagging efficiency for jets in this analysis is $33 \pm 2\%$.

In total, 115 tagged jets are selected, 60 in the electron channel and
55 in the muon channel. This compares with a Monte Carlo estimate of
69 in the electron and 45 in the muon channel. The Monte Carlo
estimate includes the Drell-Yan contribution, which has been scaled by
the factors obtained in the fit to the secondary vertex mass (see
Section \ref{sec:bfraction}), and the backgrounds listed in Section
\ref{sec:backgrounds}.  Out of the 115 tagged jets 16 are negatively tagged
compared with a Monte Carlo estimate of 16.7. One event contains two
positively tagged $b$ jets compared with a Monte Carlo expectation of
1.46.

\subsection{Backgrounds}
\label{sec:backgrounds}
Backgrounds to $Z+b$ production can arise  from misidentified
leptons,  from genuine leptons and $b$ jets coming from other
processes, or from light jets or $c$ jets
that are misidentified as $b$ jets. The first two background sources
are discussed in this section, and the latter is discussed in
section~\ref{sec:bfraction}.

The background in which one or both reconstructed electrons in the
$Z\rightarrow e^+e^-$ channel are misidentified  from other particles in the
final state is estimated from the data. The probability that a jet will pass all electron
identification criteria is determined from several inclusive jet
samples. These samples have negligible prompt electron content. The
probabilities are parameterized as a function of $E^{\rm jet}_T$ and are on
average $0.1 \%$ for central jets and $1.5 \%$ for forward
jets. Because an associated track is not required for a forward
electron, the misidentification probability is much higher
than for central electrons.

A sample of events in which exactly one trigger electron is
reconstructed is now taken. The trigger electron is paired with any
other jet in the event such that the invariant mass of the electron
and the jet
lies within the $Z$ mass window. The jet energy is taken at the electromagnetic 
calorimeter energy scale i.e. the correct energy scale for an electron or photon.
A weight, which equals the jet
misidentification probability, is assigned to the jet. The total
background to inclusive $Z$ production is then the sum of all
weights. If there is more than one jet in the event that forms an
invariant mass within the mass window of the $Z$, each combination is
used. Background distributions are derived by weighting the electron +
jets distributions with the weights.  The background for $Z$+jets
($Z+b$ jets) is the sum of those weights for events which contain at
least one jet ($b$ jet) in addition to the one paired with the
electron.

Using this method, a background contribution of $3.1 \pm 1.5 $ $b$ tagged jets is
estimated within the $Z$ mass window. 

Background in the $Z\rightarrow \mu^+\mu^-$ channel in which the electric
charges of the two muon candidates are uncorrelated  is estimated from events
with two reconstructed muon candidates that have the same electric
charge~\cite{wzinclprd}.  This background comes from events in which
one or both muon candidates arise from hadron decays or events in which hadrons
in the final state are misidentified as  muons. Due to the low statistics
of the events with a $b$ tagged jet, the fractional contribution
 of this background is
estimated as the observed ratio of generic jets (i.e., any jet
regardless of which quark flavor or gluon it originated from) with a 
like-sign $Z$ candidate to those with an unlike-sign $Z$ candidate. 
The number of $b$ tagged jets from this background source is estimated 
to be $0.24\pm 0.12$. This number is in agreement with the zero tagged 
jets with a like sign $Z$ candidate observed in the data.

The background from other processes is estimated from Monte Carlo 
simulations. The production of $t\bar{t}$ pairs is found to contribute 
$0.25 \pm 0.05 $ ($0.24 \pm 0.05 $) tagged jets to the electron (muon) 
channel. The production of $ZZ$ is found to contribute $0.36 \pm 0.07$ 
($0.28 \pm 0.06$) tagged jets to the electron (muon) channel. 
Backgrounds from other processes such as $WW$, $WZ$, $b\bar{b}$ production or
$Z\rightarrow \tau^+ \tau^-$  are estimated to be negligible.

\begin{figure}[htbp]
  \begin{center}
  \includegraphics[width=1.1\textwidth]{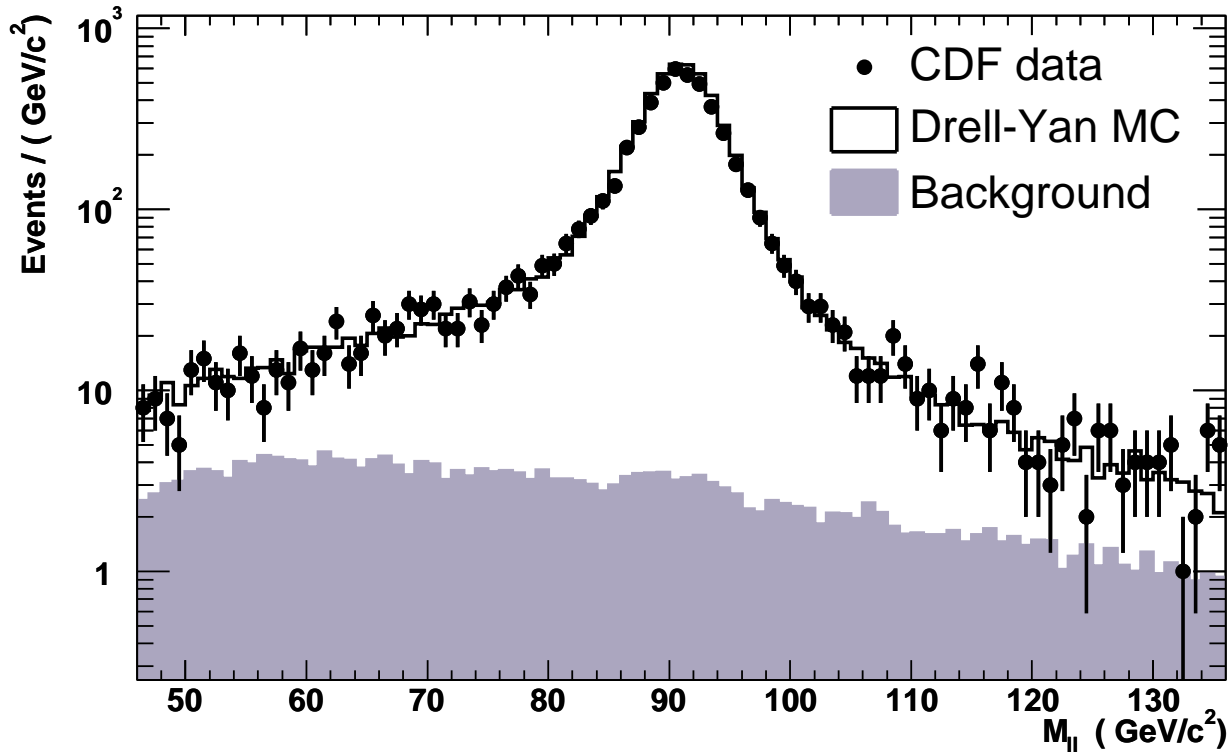}
  \caption{The invariant mass of the dilepton pair for the sample with
  jets with $E^{\rm jet}_T>20$~GeV and $|\eta^{\rm jet}|<1.5$ compared
  with the expectation from signal and background sources.  The Drell-Yan
  Monte Carlo has been normalized to the luminosity of the data sample assuming
  the NNLO Drell-Yan cross section.}
\label{fig:zmassjetsumscaled} 
  \end{center}
\end{figure}

\begin{figure}[htbp]
  \begin{center}
  \includegraphics[width=1.1\textwidth]{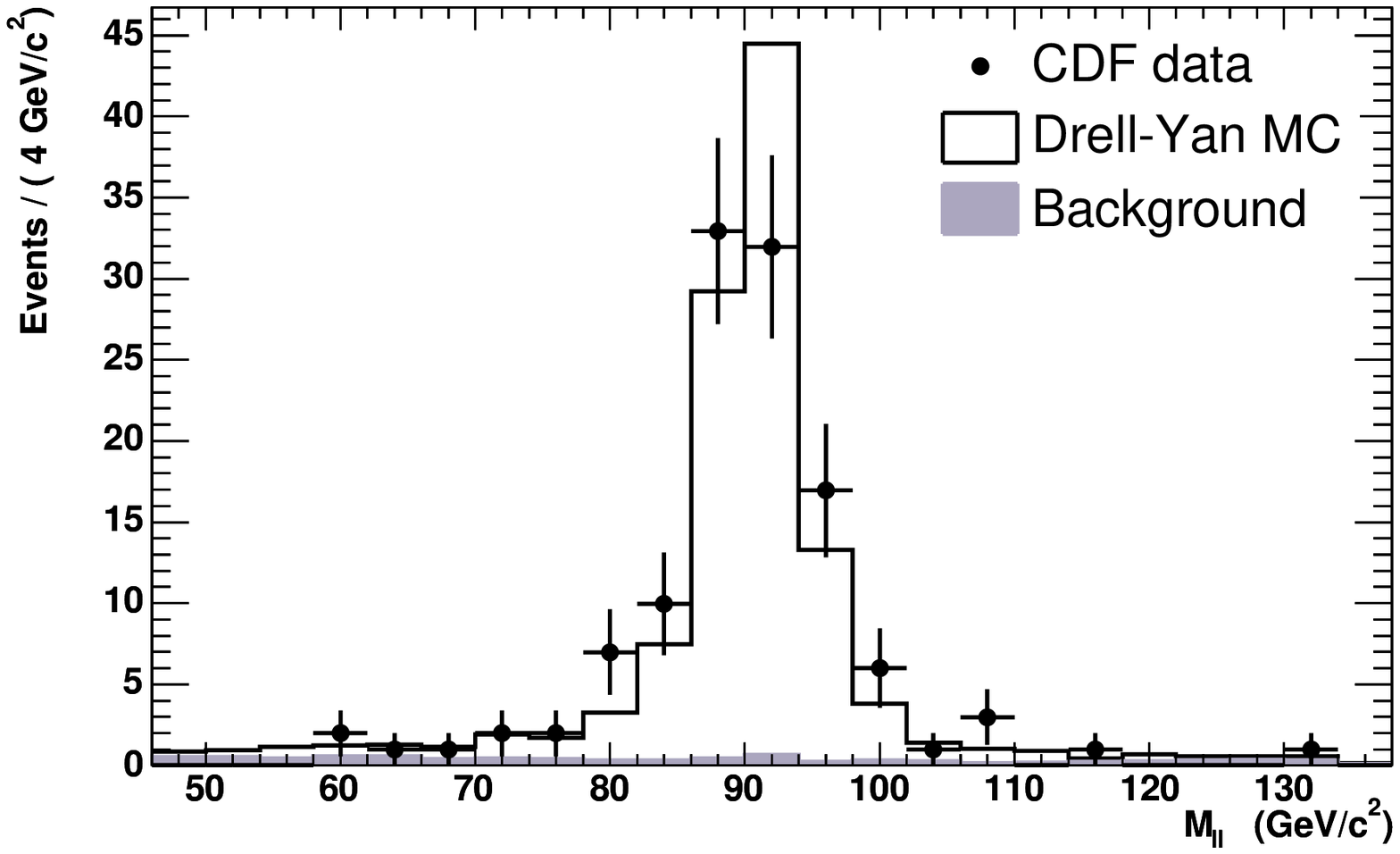}
  \caption{The invariant mass of the dilepton pair for the sample with
  positively tagged
  jets with $E^{\rm jet}_T>20$~GeV and $|\eta^{\rm jet}|<1.5$ compared
  with the expectation from signal and background sources.  The Drell-Yan
  Monte Carlo has been scaled by the factors determined in 
  Section \ref{sec:bfraction}.}
\label{fig:zmasstaggedjetsumscaled} 
  \end{center}
\end{figure}

The invariant mass of the dilepton pair is shown in
Figure~\ref{fig:zmassjetsumscaled} for events with at least one generic jet 
with $E^{\rm jet}_T>20$~GeV and $|\eta^{\rm jet}|<1.5$. The data are compared with 
the expected Drell-Yan and estimated background contributions. Good agreement is
observed near $M_{Z}$ and in the tails.
Figure~\ref{fig:zmasstaggedjetsumscaled} shows the dilepton invariant mass 
for events with at least one positively tagged jet
in the same kinematic range. The data are well modeled by the simulation, 
and a clear $Z$ signal is also observed in this sample.

\section{\boldmath Fraction of $b$ Jets}
\label{sec:bfraction}
The fraction of $b$ jets in the tagged jet sample is estimated by
performing a fit to the invariant mass of all charged tracks
attached to the secondary vertex $M_S$~\cite{msecvtx}. 
On average $b$ jets have a larger $M_S$  than $c$ jets or 
light  jets due to the larger mass of $b$ hadrons. 
The $M_S$ distribution for both positively and
negatively tagged jets is used, in order to better discriminate between
heavy and light  jets, since jets with a genuine secondary vertex
overwhelmingly have positive tags, whereas jets with a false
secondary vertex may have positive or negative tags. In addition
 using the negative tags in the fit allows
a better separation of the charm and light quark contributions that
have similar $M_S$ distribution for positively tagged jets.

\begin{figure}[htbp]
  \begin{center}
	\vspace{-1cm}
  \includegraphics[width=1.0\textwidth]{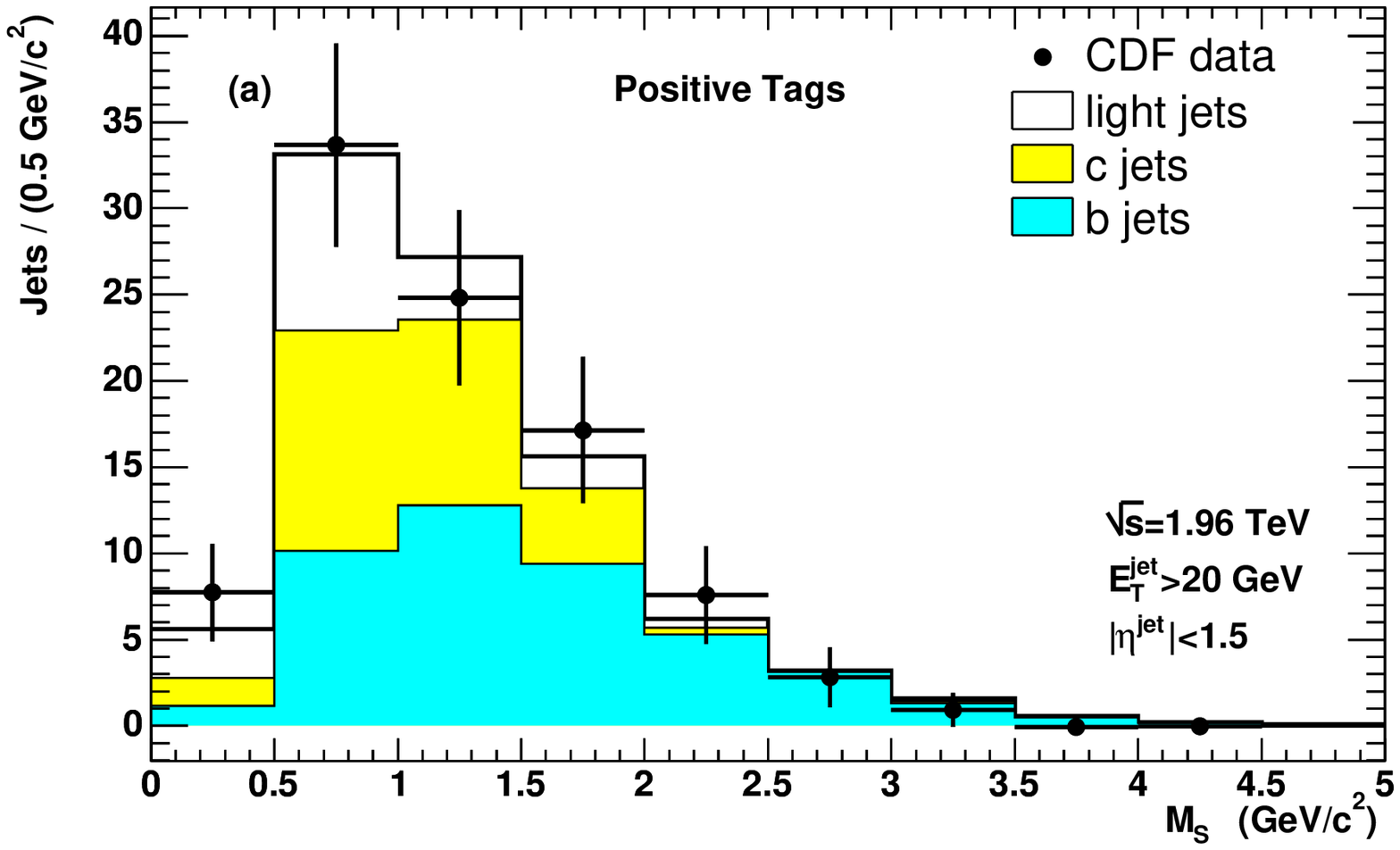}
  \includegraphics[width=1.0\textwidth]{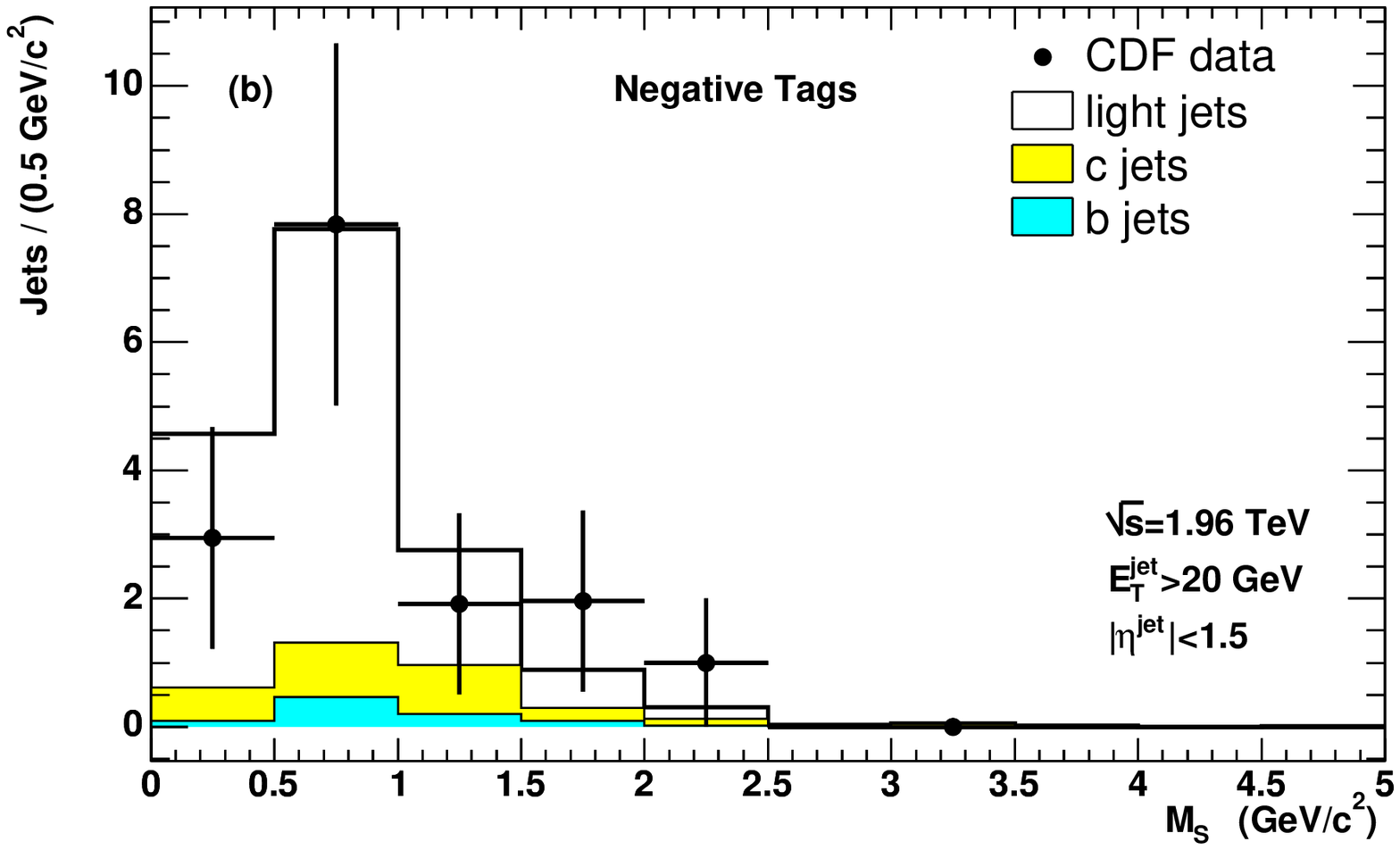}
  \caption{The mass at the secondary vertex, $M_S$, for
  (a) positively and (b) negatively tagged jets with $E^{\rm jet}_T>20$~GeV and $|\eta^{\rm
  jet}|<1.5$. The non-Drell-Yan background has been subtracted from
  the data. The data are compared with the sum of the light, $c$ and $b$ Monte Carlo templates
	after being scaled by the factors $\rho_l$, $\rho_c$, and $\rho_b$, respectively. 
  The open white area represents the light quark template, the lightly shaded are
  the $c$ quark template and the dark shaded are the $b$ quark template.}
\label{fig:massatsecvertexsumscaled} 
  \end{center}
\end{figure}

The $M_S$ distributions for positively and negatively tagged data jets are shown in
Figure~\ref{fig:massatsecvertexsumscaled}. The distributions of
$M_S$ for $b$, $c$, and light jet events are taken from
the Monte Carlo simulation. The Monte Carlo light jet distribution has 
been corrected using data as described below.

It is important to ensure that the simulation models the $M_S$ 
distributions well.  The $M_S$ distributions are affected by
the tracking efficiency,  the charged particle multiplicity 
of the $b$ hadron decay, and the fraction  of jets with 
two $b$  or $c$ hadrons. Systematic uncertainties have been
assigned to each of these contributions (see Section \ref{sec:sys}),
and the $M_S$ distribution in a Monte Carlo simulation of $b$
jets has been compared with the dijet sample used to determined the $b$
efficiency (see Section \ref{sec:selection}). The simulation is found to
describe the data within the uncertainties quoted~\cite{monica}.

The number of positive tags in light quark
 jets is larger than the number of negative tags due to long lived
 particles, such as $K^0_S$ and $\Lambda$, and due to vertices produced by
 nuclear interactions of particles with material in their
 path~\cite{secvtx}. The current 
simulation may not describe these effects accurately. In order to investigate
these effects, a sample of approximately 13,000 $\gamma + {\rm jet}$ data
events is taken, with jets in the same kinematic range as for the
$Z+b$ jets selection. This sample
was chosen since it has higher statistics than and a similar event topology
as the $Z+b$ jets sample. The  $M_S$ distribution of the $\gamma + {\rm jet}$ data  is 
obtained separately for positive and negative tags.  
The fraction of
light, $c$, and $b$ jets in each sample is determined  using Monte
Carlo distributions as templates and performing a likelihood fit, similar to
the one described below,
in which the normalization of each Monte Carlo template is allowed to
vary. It is found that the ratio of positive to negative tags for
light jets in the data is $1.49$, compared with the expectation from
the Monte Carlo of $1.93$. To correct for the difference between the data 
and the Monte Carlo simulation, the Monte
Carlo light negative tagged jets are increased by a factor of
$1.93/1.49=1.30$. This reweighted template shape is taken for the central value,
and the full difference from unity, $\pm 30\%$, is taken as the systematic 
uncertainty. 

A binned maximum likelihood fit using Poisson statistics is performed
to the $M_S$ distributions of the positively and negatively tagged jets in
the $Z$ sample for range $M_S<3.5 (1) $~GeV/$c^2$ for positively
(negatively) tagged jets. The range is chosen in order to have enough
statistics for the fit. The Monte Carlo distributions are taken as
templates and scaled by factors $\rho_b$, $\rho_c$, and $\rho_l$ for
$b$, $c$, and light jets, respectively. The data distribution is used
after subtraction of the non-Drell-Yan background as estimated in
Section
\ref{sec:backgrounds}.  The quantity $\rho_b$ is thus the number of
fitted reconstructed signal $b$ jets in data divided by the number
 in the simulation.  Only statistical errors from the data and
Monte Carlo are used in the fit. The fit takes into account the Monte
Carlo statistical errors using the method described in
Ref.~\cite{tfractionfitter}. The fit gives values of $\rho_b=0.93 \pm
0.29$, $\rho_c=1.69
\pm 0.94$ and $\rho_l= 1.36 \pm 0.53$. The correlation coefficient between
$\rho_b$ and $\rho_c$ is $-0.68$ and between $\rho_b$ and $\rho_l$ 
is $0.10$.

The number of $b$ jets in the $Z$ sample after
subtracting the background contributions is estimated to be $N_{\rm
Data}(Z+b{\,\rm jets})=45 \pm 14$.  A check is performed by fitting
only positively tagged jets, and good agreement is obtained
with a value of $\rho_b= 0.95 \pm  0.31$.

\section{Cross Section}
\label{sec:crosssection}
The inclusive $b$ jet cross section $\sigma(Z+b\, {\rm jets})$ is
proportional to the number of $b$ jets with $E^{\rm jet}_T>20$~GeV and
$|\eta^{\rm jet}|<1.5$ and is defined for dilepton masses
$66<\mll<116$~GeV$/c^2$. The branching fraction ${\cal B}(Z\rightarrow
\ell^+\ell^-)$ is defined for a single lepton flavor.
All the cross sections and cross section ratios presented 
are fully corrected for detector response and resolution, and 
presented at the hadron level. No corrections are made for the 
underlying event or for hadronization effects (see Section \ref{sec:selection}).
They are defined for jets with a cone size of 0.7 that include a b hadron within
this cone. 

A ratio method is used to extract the cross section. In doing so use can be made
of the uncertainties estimated for the inclusive $Z$ cross section
measurement~\cite{wzinclprd} for the lepton and trigger selection. 
First a measurement is made of the ratio of the
$Z+ b$ jet cross section to the total $Z$ cross section:

\begin{equation}
\frac{\sigma(Z+b {\,\rm jets})}{\sigma (Z)} = 
\frac{N_{\rm Data}(Z+b{\,\rm jets})/\epsilon(Z+b{\,\rm jets})}
{N_{\rm Data}(Z)/\epsilon(Z)},
\end{equation}

\noindent where $N_{\rm Data}(Z+b{\,\rm jets})$ is the fitted number
of $b$ jets (see Section \ref{sec:bfraction}) and $N_{\rm Data}(Z)$ is the total number of events 
with a lepton pair in the mass range 
$66<\mll<116$~GeV/c$^2$ in the data. 
In both cases the number of data events is taken after subtraction of 
the background contributions (see Section \ref{sec:backgrounds}). The efficiencies of 
the $Z+b{\,\rm jet}$ and the $Z$ boson selections are 
 $\epsilon(Z+b{\,\rm jet})=7.7\%$ and $\epsilon(Z)=27\%$, respectively. 
They are determined from {\sc pythia} Monte Carlo simulation and are
corrected for any differences from the data. The ratio
$\epsilon(Z+b{\,\rm jets})/\epsilon(Z)$ is also determined using {\sc herwig} v6.5 \cite{herwig} and 
a similar result is obtained (0.286 for {\sc herwig} compared with 0.285 for {\sc pythia}). 

The cross section is then calculated as:

\begin{eqnarray}
\sigma(Z+b {\, \rm jets})\times {\cal B}(Z\rightarrow \ell^+\ell^-) 
= \frac{\sigma(Z+b {\,\rm jets})}{\sigma (Z)} 
\cdot \sigma_{\rm CDF}(Z)\times {\cal B}(Z\rightarrow \ell^+\ell^-),
\end{eqnarray}

\noindent where $\sigma_{\rm CDF}(Z)\times {\cal B}(
Z\rightarrow \ell^+\ell^-)=254.9 \pm 3.3 \mbox{(stat.)} \pm 4.6 \mbox{(syst.)} \pm
15.2 \mbox{(lum.)}$~pb is the CDF measurement of the $Z$ production cross
section times branching
fraction  for
 a single lepton flavor~\cite{wzinclprl}.

A measurement is also made of the ratio of the $Z+ b$ jets cross
section to the $Z+$ generic jets cross section in order 
to measure the fraction of jets that contain
at least one $b$ hadron. The $Z+$ generic jets cross section is proportional
to the number of generic jets with $E^{\rm jet}_T>20$~GeV and $|\eta^{\rm jet}|<1.5$. The ratio of the $Z+b$ jets to $Z +$ generic jets cross section 
is obtained from the data as:

\begin{eqnarray}
 \frac{\sigma(Z+b {\,\rm jets})}{\sigma (Z+ {\,\rm jets})} =
\frac{N_{\rm Data}(Z+b{\,\rm jets})/\epsilon(Z+b{\,\rm jets})}
{N_{\rm Data}(Z+{\rm jets})/\epsilon(Z+{\rm jets})},
\end{eqnarray}

\noindent where $N_{\rm Data}(Z+\rm jets)$ is the number of generic data 
jets for events with a lepton pair within the $Z$ mass
window after background subtraction and $\epsilon(Z+\rm jets)=24\%$ is 
the efficiency of the $Z+\rm jet$ selection. 

\begin{figure}[htbp]
  \begin{center}
  \includegraphics[width=1.07\textwidth]{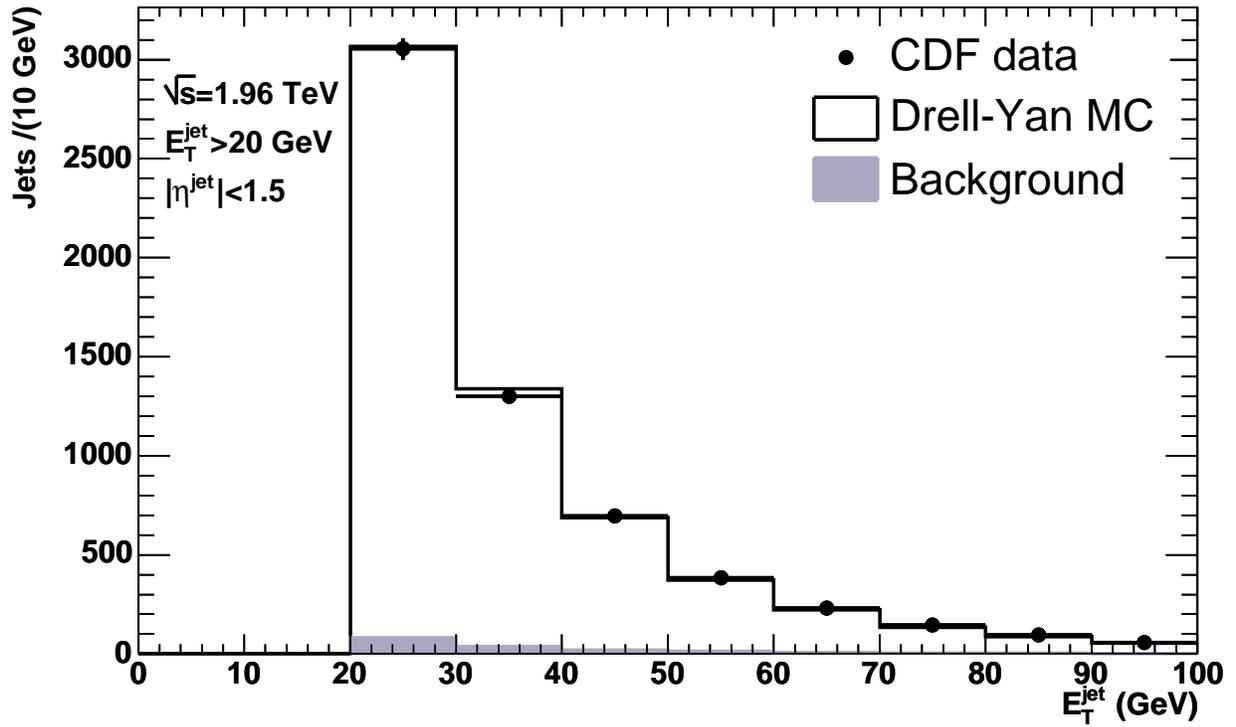}
  \includegraphics[width=1.07\textwidth]{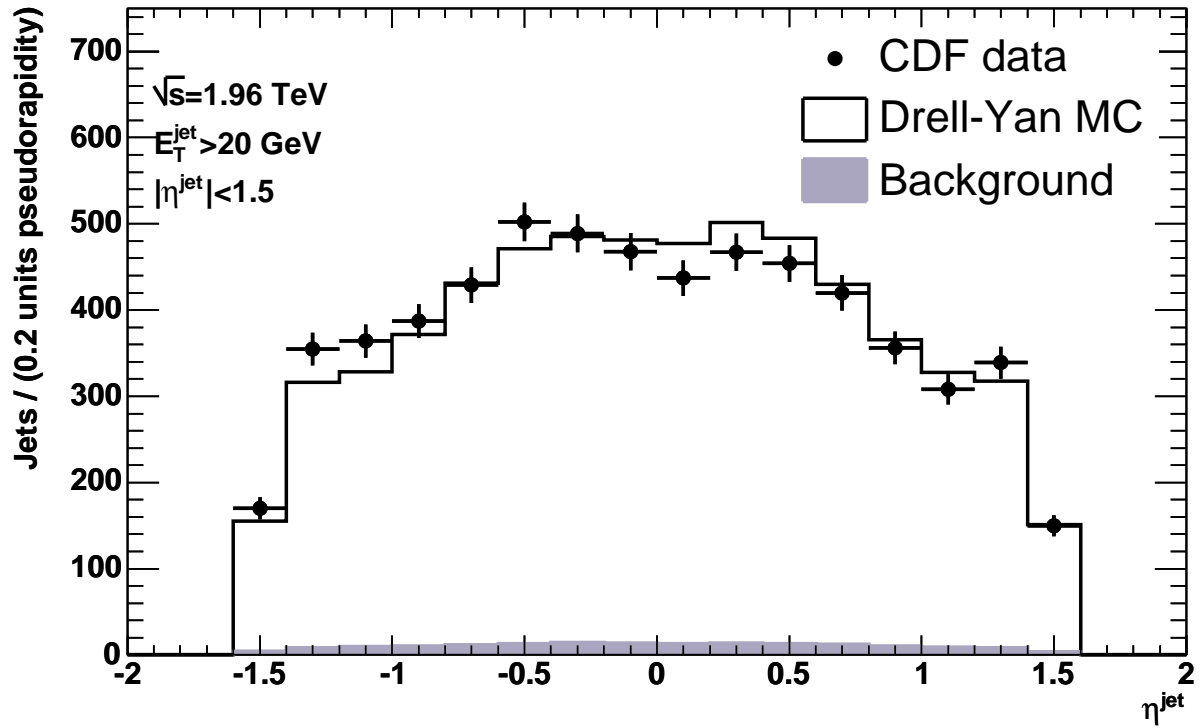}
  \caption{The $E^{\rm jet}_T$ and $\eta^{\rm jet}$ distributions for
  generic jets with $E^{\rm jet}_T>20$~GeV and $|\eta^{\rm
  jet}|<1.5$. The Drell-Yan Monte Carlo has been scaled such that the
  total number of jets in the simulation is the same as in the data.}
  \label{fig:genericjet} \end{center}
\end{figure}

\begin{figure}[htbp]
	\vspace{-0.5cm}
  \begin{center}
  \includegraphics[width=1.0\textwidth]{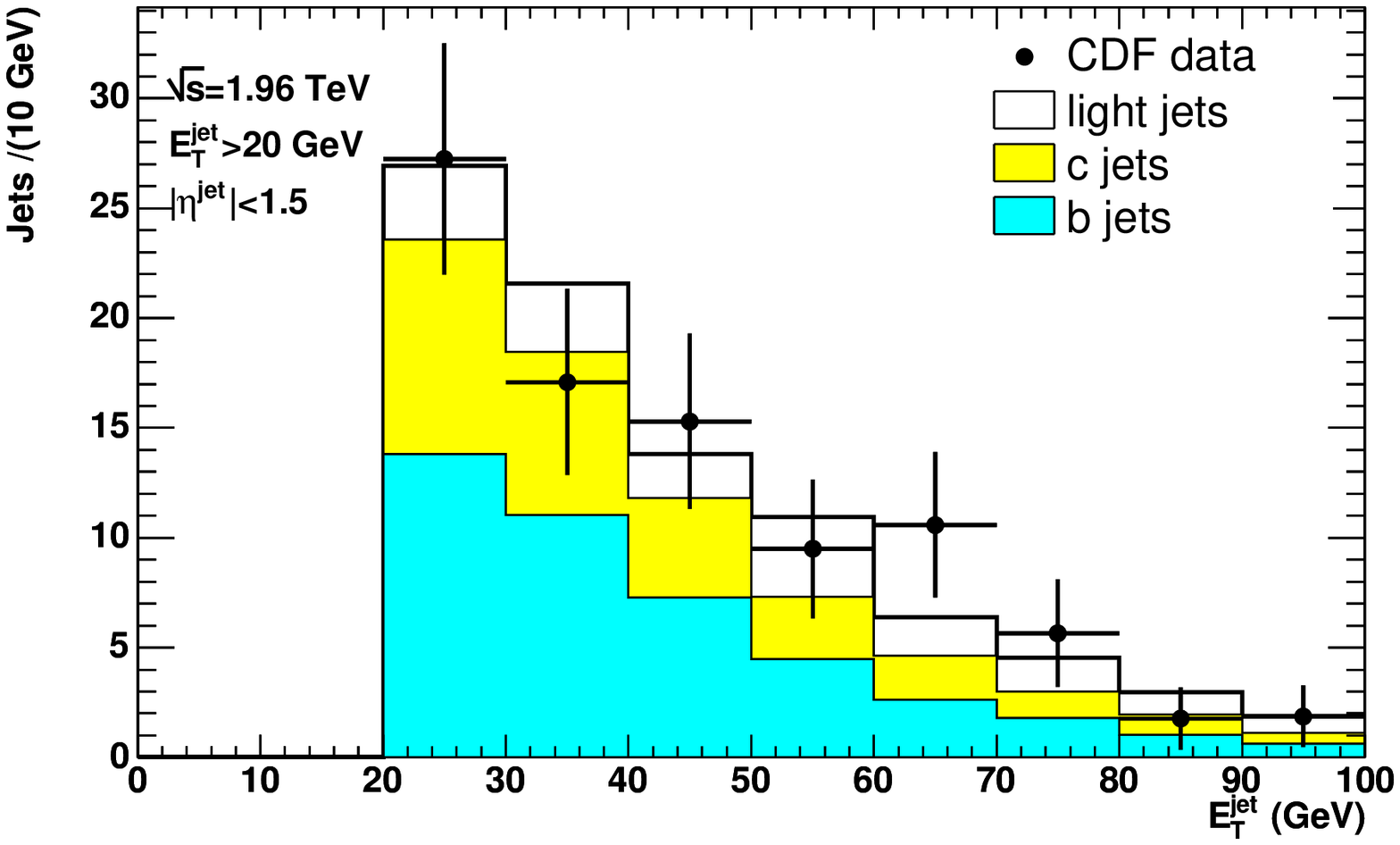}
  \includegraphics[width=1.0\textwidth]{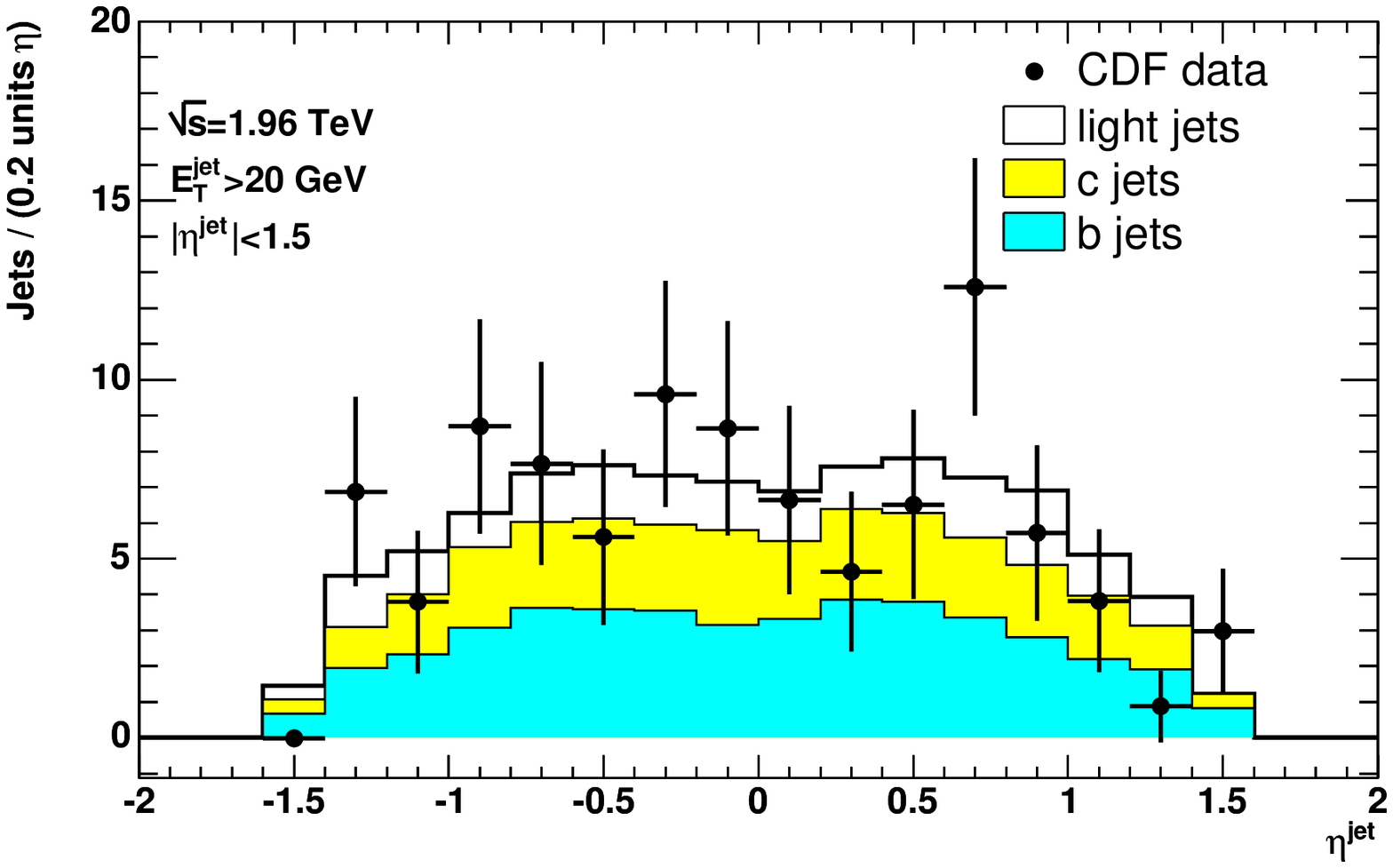}

  \caption{The $E^{\rm jet}_T$ and $\eta^{\rm jet}$ distributions for
  positively tagged jets with $E^{\rm jet}_T>20$~GeV and $|\eta^{\rm
  jet}|<1.5$. The non-Drell-Yan background has been subtracted from
  the data. The data are compared with the sum of the light, $c$ and $b$ 
  contributions after being scaled by the factors $\rho_l$, $\rho_c$, and $\rho_b$, respectively
 (see Section
  \ref{sec:bfraction}). The open white area represents the light quark 
  template, the lightly shaded are
  the $c$ quark template and the dark shaded are the $b$ quark template.}
\label{fig:bjet} 
  \end{center}
\end{figure}

The validity of the kinematic distributions in the Monte Carlo simulation
was checked by comparing the transverse energy and pseudorapidity distributions
for generic jets and for the positively $b$ tagged jets. 
These distributions, which are shown in
Figure \ref{fig:genericjet} for generic jets and in Figure \ref{fig:bjet} for $b$ jets, demonstrate
that {\sc pythia} describes the data well and thus  may be used to correct the data.

\section{Systematic Uncertainties}
\label{sec:sys}
The total systematic uncertainty on the measurement is estimated by adding
the uncertainties described in this section in quadrature.

  The uncertainty on the jet energy scale was estimated by changing 
the scale in the Monte Carlo simulation by the uncertainty following
the procedure of Ref.~\cite{jes_nim}.  An additional uncertainty of
$\pm 0.6\%$~\cite{topmassprd} is applied on the $b$ jets to account
for possible differences in the description of $b$ jets by the Monte Carlo
simulation compared with light jets.

Possible differences between the Monte Carlo simulation and data
$E^{\rm jet}_T$ and $\eta^{\rm jet}$ distributions, which may
arise for many reasons including uncertainties in the parton distribution
functions, inadequacies in the QCD model  or uncertainties
in the $b$ quark fragmentation, are accounted for
in the systematic error by weighting the Monte Carlo hadron level
distributions by $(E_{T\rm Had}^{\rm jet}/35)^{\pm \alpha_{E_T}}$ and
$(|\eta_{\rm Had}^{\rm jet}|+0.5)^{\pm \alpha_\eta}$, respectively. Here
$E_{T\rm Had}^{\rm jet}$ is the transverse energy in GeV, and
$\eta_{\rm Had}^{\rm jet}$ is the pseudorapidity of the Monte Carlo
hadronic jet. The values of 35 GeV and 0.5 units of pseudorapidity are
arbitrary and chosen to be close to the center of the distributions. 
The parameter $ \alpha_{E_T}$ is set to 0.2 (1.0) for
untagged (tagged) jets. The parameter $ \alpha_{\eta}$ is set to 0.2
(1.0) for untagged (tagged) jets. These parameters are chosen by
comparing the weighted and unweighted Monte Carlo reconstructed
distributions with those of the data and choosing the weight such that it is still
statistically compatible with the data.

The $b$ jet tagging efficiency has a $6.6\%$ uncertainty~\cite{secvtx}.

Since any changes in the shapes of the light, $c$ and $b$ jet
templates can change the value of $\rho_b$
obtained, various possible variations of the template shapes are
included in the systematic uncertainty. An uncertainty in the negative to
positive light jet tag rate is estimated by weighting the negative tags
of the light template by $\pm 30\%$.  This is the
observed difference between data and Monte Carlo simulation for the
negative to positive light jet tag ratio (see
section~\ref{sec:bfraction}). Since it is not {\it a priori} known whether
the jets contain one or two heavy quarks, an uncertainty in the shape
of the $c$ ($b$) quark template is estimated by taking half the
difference between results obtained using the templates when there is one or 
two $c$ ($b$)
hadrons within a cone of 0.7 around the jet axis. An uncertainty 
due to possible track reconstruction inefficiencies is estimated by
recalculating templates after randomly removing $3\%$ of the tracks 
in the Monte Carlo simulation.
 An uncertainty in the shape of the $b$ quark template
due to the uncertainty in the mean $b$ hadron charged particle
multiplicity is estimated by weighting the Monte Carlo multiplicity
such that the mean changes by $\pm 0.15$, which is the difference
between the Monte Carlo simulation used in this analysis and data
measurements~\cite{lepjetmulti}. This systematic uncertainty also covers
any differences observed between data and simulation in the $b$-quark
rich sample used to determine the efficiency  (see Section \ref{sec:selection}). 

An uncertainty on the misidentified electron and  muon backgrounds is estimated
to be $\pm50\%$ of the estimated contribution (see
Section~\ref{sec:backgrounds}). An uncertainty in the $ZZ$ and $t\bar{t}$ backgrounds 
of $\pm20\%$ is estimated. These uncertainties include the NLO cross
section uncertainty estimated by varying the factorization scale by a factor two~\cite{ttbar,zz}, 
the cross section uncertainty due to the uncertainty on the top quark mass,
and experimental uncertainties (e.g. the lepton identification and $b$ jet tagging efficiencies).

 A 1.8\% uncertainty on the selection efficiency of $Z$ events is taken
from Ref.~\cite{wzinclprl}. A 5.8\% uncertainty is taken for the luminosity 
determination \cite{lumerr}.

The effect of these contributions on the systematic uncertainty on the cross
section is listed in Table~\ref{tab:systable}. The systematic uncertainties
are evaluated separately for the cross section ratio measurements and
are also included in Table~\ref{tab:systable}.
The total systematic uncertainty on the cross section is 22\%
compared with a statistical precision of 31\%. The largest uncertainty arises
from the assumed Monte Carlo $E^{\rm jet}_T$ distribution and will be reduced with 
higher data statistics to constrain this distribution.
\begin{table}[h]
\caption{\label{tab:systable}
The systematic uncertainties on the cross section
and ratio measurements. The total systematic uncertainty on
each measurement is estimated by adding the individual uncertainties in
quadrature.} 
\begin{center}
\begin{tabular}{|l|c|c|c|} \hline
Source of Uncertainty & $\delta (\sigma  (Z+b {\, \rm
jets})) $   & $\delta (\sigma  (Z+b {\, \rm
jets})/\sigma  (Z))$ & $\delta (\sigma  (Z+b {\, \rm
jets})/\sigma  (Z+ {\rm jets}))$  \\ 
 & (\%) &  (\%) &  (\%) \\ \hline 
jet energy scale               &  $3.9$& $3.9$ & $6.7$   \\ 
$b$ jet energy scale           &  $0.2$& $0.2$ & $0.2$  \\ 
MC $\eta^{\rm jet}$ dependence &  $6.6$& $6.6$ &$6.6$ \\ 
MC $E^{\rm jet}_T$ dependence  & $16.1$& $16.1$ & $16.2$  \\
$b$ tagging efficiency         &  $6.6$& $6.6$ & $6.6$  \\
light jet template             &  $1.8$& $1.8$ & $1.8$ \\
single/double $c$ quark  in jet &  $2.8$& $2.8$ & $2.8$ \\
single/double $b$ quark  in jet &  $3.6$& $3.6$ & $3.6$ \\
track reconstruction efficiency &  $9.9$& $9.9$ & $9.9$ \\
$b$ hadron multiplicity   & $1.0$& $1.0$ & $1.0$  \\
misidentified lepton background    & $1.3$& $0.9$ & $0.4$ \\
other backgrounds         & $0.3$& $0.3$ & $0.3$  \\ 
$Z$ selection efficiency  & $1.8$& n. a.  & n. a. \\
luminosity  &$5.8$& n. a. & n. a.  \\ \hline
total       &$22.9$& $22.1$ & $22.8$ \\ \hline
\end{tabular}
\end{center}

\end{table}

\section{Results}
\label{sec:result}
The $b$ jet cross section for events with $66<\mll<116$~${\rm GeV}/c^2$ and $E^{\rm
jet}_T>20$~${\rm GeV}$ and $|\eta^{\rm jet}|<1.5$ is measured to be

$$\sigma (Z+b {\,\rm jets})\times {\cal B}(Z \to \ell^+ \ell^-) = 0.93 \pm 0.29 \mbox{(stat.)} \pm 0.21 \mbox{(syst.)} \mbox{ pb.}$$ 
The theoretical prediction of $0.45 \pm 0.07$~pb is consistent with the
measurement. 
The theoretical cross section and the ratios listed below were evaluated at NLO using 
{\sc MCFM}~\cite{zbmcfm,campbellprivate} with the parton distribution functions from 
CTEQ6M~\cite{cteq6}. The factorization and renormalization scales were set to
 $M_{Z}$. The uncertainty on the NLO prediction includes the uncertainty arising from the
choice of these scales, which were varied from $M_Z/2$ to $2M_Z$, the 
uncertainty on the parton distribution functions and the uncertainty 
on the strong coupling constant.

The ratio of $Z+b {\,\rm jets}$ production to inclusive $Z$ production is measured to be
$$\sigma(Z+b {\,\rm jets})/\sigma (Z)=0.0037 \pm 0.0011 \mbox{(stat.)}  \pm 
0.0008  \mbox{(syst.)}.$$  
The {\sc pythia}
estimate of 0.0035 and the NLO QCD calculation of $0.0019 \pm 0.0003$
both agree with the measured value.

The NLO QCD cross section $\sigma (Z+b {\,\rm jets})$ and the ratio
$\sigma(Z+b {\,\rm jets})/\sigma (Z)$ do not include the effects of underlying
event and hadronization. These effects are estimated from {\sc pythia} to
change the NLO QCD cross section and the ratio to the inclusive $Z$ cross section
by $+10\%$ for the underlying event and $-1\%$ for the hadronization.

The ratio of $Z+b {\,\rm jets}$ production compared to $Z+ {\,\rm jets}$
production is 
measured to be 
$$\sigma(Z+b {\,\rm jets})/\sigma (Z+ {\,\rm jets})=0.0236 \pm 0.0074  \mbox{(stat.)}
 \pm 0.0053   \mbox{(syst.)} $$  
The estimate from {\sc pythia}  of  $0.0218$ and the
NLO QCD calculation of $0.0181 \pm 0.0027$ agree with this result.  The NLO QCD cross section has
not been corrected for underlying event and hadronization. These
effects are estimated from {\sc pythia} and change the value
by $-7\%$ and $+1\%$, respectively. 
The  \Dzero collaboration~\cite{zbjetd0} obtained a measurement of
$0.021 \pm 0.004 \mbox{(stat.)} ^{+0.002}_{-0.003} \mbox{(syst.)}  $
for the fraction of events  with at least one $b$ jet, with  $P^{\rm
jet}_T>20$~${\rm GeV}/c$ and $|\eta^{\rm jet}|<2.5$, to those events
with at least one generic jet within the same kinematic range.
Note that in the analysis by the \Dzero collaboration, an NLO QCD prediction for the
ratio of the $Z+b$ quark to the $Z+c$ quark production rate is assumed. 
This assumption results in smaller uncertainties for the \Dzero analysis despite the 
present analysis having more than twice the number of tagged
$b$ jets and similar $b$ purity. The present analysis does not take the approach 
followed by the \Dzero collaboration in order to maintain independence of the 
measurement from QCD calculations. 

\section{Conclusions}
\label{sec:conclusion}
In conclusion, the inclusive $b {\,\rm jets}$ production cross section
in events with a $Z$ boson has been measured for jets with
$E_T>20$~GeV and $|\eta^{\rm jet}|<1.5$ in $p\bar{p}$ collisions at
$\sqrt{s}=1.96$~TeV at the Tevatron using CDF data with an integrated
luminosity of about 330 pb$^{-1}$.

The $Z+b {\,\rm jets}$ cross section times branching ratio is measured as $0.93 \pm 0.36$~pb
for the $Z$ boson decaying into a single charged lepton flavor. The ratio of the $Z+b {\,\rm jets}$ to the $Z+{\,\rm jets}$ cross
section is $2.36 \pm 0.92\%$.
This is the first measurement of the $Z+b {\,\rm jets}$ cross section
without any assumptions on the $Z+c {\,\rm jets}$ production rate.
The uncertainty is dominated by the limited statistical 
precision, which is expected to improve significantly in the near future. 

Within current uncertainties, the theoretical predictions agree with
these measurements, which are sensitive to the $b$ quark density at high
momentum transfer.

\begin{center}
\textbf{Acknowledgments}
\end{center}

We thank the Fermilab staff and the technical staffs of the participating
institutions for their vital contributions.  We also thank John Campbell
for his help on the theoretical prediction.

This work was supported by the U.S. Department of Energy 
and National Science Foundation; the Italian Istituto Nazionale di Fisica Nucleare; the 
Ministry of Education, Culture, Sports, Science and Technology of Japan; the Natural 
Sciences and Engineering Research Council of Canada; the National Science Council of 
the Republic of China; the Swiss National Science Foundation; the A.P. Sloan Foundation; 
the Bundesministerium f\"ur Bildung und Forschung, Germany; the Korean Science and 
Engineering Foundation and the Korean Research Foundation; the Particle Physics and 
Astronomy Research Council and the Royal Society, UK; the Russian Foundation for Basic 
Research; the Comisi\'on Interministerial de Ciencia y Tecnolog\'{\i}a, Spain; in part 
by the European Community's Human Potential Programme under contract HPRN-CT-2002-00292; 
and the Academy of Finland.


\begin{thebibliography}{99} 
\bibitem{zbmcfm} J. Campbell, R.K. Ellis, F. Maltoni and S. Willenbrock, Phys. Rev. D~\textbf{69}, 074021 (2004).


\bibitem{bbh}
D. Dicus, T. Stelzer, Z. Sullivan and S. Willenbrock, Phys. Rev. D~\textbf{59}, 094016 (1999);
C. Balazs, H. J. He and C. P. Yuan, Phys. Rev. D~\textbf{60}, 114001 (1999);
F. Maltoni, Z. Sullivan and S. Willenbrock, Phys. Rev. D~\textbf{67}, 093005 (2003).

\bibitem{mssmhiggs} D. Choudhury, A. Datta and S. Raychoudhuri, arXiv:hep-ph/9812201; 
C. S. Huang and S. H. Zhu, Phys. Rev. D~\textbf{60}, 075012 (1999);
J. Campbell, R.K. Ellis, F. Maltoni and S. Willenbrock, Phys. Rev. D~\textbf{67}, 095002 (2003).


\bibitem{singletop} T. Stelzer, Z. Sullivan and S. Willenbrock, Phys. Rev. D~\textbf{56}, 5919 (1997);
T. Stelzer, Z. Sullivan and S. Willenbrock, Phys. Rev. D~\textbf{58}, 094021 (1998).

\bibitem{smhiggs} M. Carena {\it et al.}, Higgs Working Group Collaboration, arXiv:hep-ph/0010338.

\bibitem{wzinclprl} 
D. Acosta \textit{et al}. (CDF  Collaboration),
Phys.\ Rev.\ Lett.~{\bf 94}, 091803 (2005).

\bibitem{wzinclprd} 
A. Abulencia \textit{et al}. (CDF  Collaboration), 
arXiv:hep-ex/0508029, submitted to Phys. Rev. D.

\bibitem{d0wz} B. Abbott {\it et al.} (\Dzero Collaboration), Phys. Rev. D~\textbf{60}, 052003 (1999).

\bibitem{zjets}  F. Abe {\it et al.} (CDF Collaboration), Phys.\ Rev.\ Lett.~\textbf{77}, 448  (1996).

\bibitem{CDF_run2} D. Acosta \textit{et al.} (CDF Collaboration), 
Phys. Rev. D~\textbf{71}, 032001 (2005).

\bibitem{zbjetd0}  V. M. Abazov {\it et al.} (\Dzero Collaboration), Phys. Rev. Lett.~\textbf{94}, 161801 (2005).

\bibitem{collins} M. A. G. Aivazis, J. C. Collins, F. I. Olness and W. K. Tung, Phys. Rev. D~\textbf{50}, 3102 (1994); 
J. C. Collins, Phys. Rev. D~\textbf{58}, 094002 (1998); 
J. Pumplin  {\it et al.}, JHEP~\textbf{0207}, 012 (2002).

\bibitem{h1} A. Aktas {\it et al.} (H1 Collaboration), Eur. Phys. J. C~\textbf{45},  23 (2006);
A. Aktas et al., Eur. Phys. J. C~\textbf{40},  349 (2005).

\bibitem{L00} C. S. Hill \textit{et al}., Nucl. Instrum. Methods A~\textbf{530}, 1 (2004).

\bibitem{SVX} A. Sill \textit{et al}., Nucl. Instrum. Methods A~\textbf{447}, 1 (2000).

\bibitem{ISL} A. Affolder \textit{et al}., Nucl. Instrum. Methods A~\textbf{453}, 84 (2000).

\bibitem{COT} A. Affolder \textit{et al}., Nucl. Instrum. Methods A~\textbf{526}, 249 (2004).

\bibitem{cdfjpsi}  D. Acosta {\it et al.} (CDF Collaboration), Phys. Rev. D~\textbf{71}, 032001 (2005).

\bibitem{cem} L. Balka \textit{et al.}, Nucl. Instrum. Methods A~\textbf{267}, 272 (1988).

\bibitem{cha}  S. Bertolucci \textit{et al.}, Nucl. Instrum. Methods A~\textbf{267}, 301 (1988).

\bibitem{pem}  M. Albrow \textit{et al.}, Nucl. Instrum. Methods A~\textbf{480}, 524 (2002).

\bibitem{jes_nim}
A. Bhatti \textit{et al}., arXiv:hep-ex/0510047, submitted to 
 Nucl. Instrum. Methods.

\bibitem{cmu}  G. Ascoli \textit{et al.}, Nucl. Instrum. Methods A~\textbf{268}, 33 (1988).

\bibitem{CLC} D. Acosta  \textit{et al.}, Nucl. Instrum. Methods A~\textbf{461}, 540 (2001).

\bibitem{pythia} 
T.~Sj\"{o}strand and S.~Mrenna, LU-TP 01, arXiv:hep-ph/0108264.

\bibitem{cteq5l} 
H.L.~Lai \textit{et al}., Eur. Phys. J. C~\textbf{12}, 375 (2000).

\bibitem{ue} 
D. Acosta \textit{et al.} (CDF  Collaboration), Phys. Rev. D~\textbf{70}, 072002 (2004).

\bibitem{qq}
P. Avery, H. Read and G. Trahern, CLEO Report CSN-212 (1985), unpublished.

\bibitem{geant} GEANT, Detector description and simulation tool,
CERN Program Library Long Writeup W5013 (1993).

\bibitem{cdfsim} E. Gerchtein and M. Paulini, Computing in High Energy and 
Nuclear Physics, preprint arXiv:physics/0306031, (2003).


\bibitem{znnlo}
P.J. Sutton, A. D. Martin, R. G. Roberts and W.J. Stirling, Phys. Rev. D~\textbf{45}, 2349 (1992); 
P.J. Rijken and W.L. van Neerven, Phys. Rev. D~\textbf{51}, 44 (1995); 
R. Hamberg, W. van Neerven and T. Matsuura, Nucl. Phys. B~\textbf{359}, 343 (1991);
R.V. Harlander and W.B. Kilgore, Phys. Rev. Lett.~\textbf{88}, 201801 (2002);
W. van Neerven and E. Zijstra, Nucl. Phys. B~\textbf{382}, 11 (1992).

\bibitem{ttbar} N. Kidonakis, R. Vogt, Eur. Phys. J. C~\textbf{33},466-S468 (2004).
\bibitem{zz} J.M. Campbell and R.K. Ellis, Phys. Rev. D~\textbf{60}, 113006 (1999). 

\bibitem{ht} F. Abe \textit{et al.} (CDF Collaboration), Phys. Rev. Lett.~\textbf{75}, 3997 (1995).

\bibitem{secvtx} 
D. Acosta \textit{et al.} (CDF Collaboration), Phys. Rev. D~\textbf{71}, 052003 (2005).

\bibitem{msecvtx}
  D.~J.~Jackson,
  Nucl.\ Instrum.\ Methods\ A~\textbf{388} (1997) 247.
  J.~Abdallah {\it et al.} (DELPHI Collaboration),
  Eur.\ Phys.\ J.\ C~\textbf{32} (2004) 185.

\bibitem{monica} M. D'Onofrio, Ph.\ D.\ thesis, University of Geneva (2006).

\bibitem{tfractionfitter} R. Barlow and C. Beeston, Comp.\ Phys.\ Comm.\ ~\textbf{77} (1993) 219.

\bibitem{herwig}
G. Corcella {\it et al.}, JHEP~\textbf{0101} (2001)  010.

\bibitem{topmassprd}
A. Abulencia \textit{et al.} (CDF Collaboration), Phys. Rev. D~\textbf{73}, 032003 (2006).


\bibitem{lepjetmulti}
LEP/SLD Heavy Flavour Working Group, D.~Abbaneo {\it et al.}, LEPHF 2001-01;

(available from
http://lepewwg.web.cern.ch/LEPEWWG/heavy/).

\bibitem{lumerr} S. Klimenko, J. Konigsberg, and T.M. Liss, FERMILAB-FN-0741 (2003).

\bibitem{campbellprivate} J. Campbell (private communication).

\bibitem{cteq6} 
J. Pumplin {\it et al.},
JHEP~\textbf{07}, 012 (2002).

\end{thebibliography}
\end{document}